\documentclass[english]{article}
\usepackage[T1]{fontenc}
\usepackage[latin9]{inputenc}
\usepackage[a4paper]{geometry}
\usepackage{babel}
\usepackage{amsthm}
\usepackage{amsmath}
\usepackage{amssymb}
\usepackage{esint}
\usepackage[unicode=true,pdfusetitle,
 bookmarks=true,bookmarksnumbered=false,bookmarksopen=false,
 breaklinks=false,pdfborder={0 0 1},backref=false,colorlinks=false]
 {hyperref}

\makeatletter
\numberwithin{equation}{section}
\numberwithin{figure}{section}

\makeatother

\begin{document}

\title{Regularization of multi-soliton form factors in sine-Gordon model}

\author{T. P\'almai\footnote{E-mail address: palmai@phy.bme.hu}\\ \, \\ {\it Department of Theoretical Physics,}\\ {\it Budapest University of Technology and Economics,}\\ {\it H-1111 Budafoki \'ut 8, Hungary}}

\maketitle

\begin{abstract}
A general and systematic regularization is developed for the exact solitonic form factors of exponential operators in the (1+1)-dimensional sine-Gordon model by analytical continuation of their integral representations. The procedure is implemented in Mathematica. Test results are shown for four- and six-soliton form factors.
\end{abstract}

\section{Introduction}

Form factors (matrix elements of local operators) are important quantities
in quantum field theories. It is a remarkable feature
of certain two-dimensional field theories (integrable models), that
their S-matrices can be obtained exactly in the framework of factorized
scattering theory \cite{Zamolodchikov:1978,Mussardo:1992}. Furthermore, in
integrable models there is a rather restrictive set of equations satisfied
by the form factors (that is the form factor axioms \cite{Smirnov:1992,Mussardo:1992}),
which makes it possible in many cases to obtain them exactly as well.
For instance, in the sine-Gordon model all form factors of exponential
operators are known \cite{Smirnov:1992,Lukyanov:1997,Babujian:1998}. The spectrum
of the sine-Gordon model consists of a soliton-antisoliton doublet
and their bound states, called {}``breathers''. While
the breather-breather form factors can be given explicitly (see e.g.
\cite{Lukyanov:1997}), the solitonic ones, in general, are only known
in terms of some highly non-trivial integral representations. In addition,
the integrals converge in a limited domain of the parameters. In this
paper we give a regularization procedure to calculate the solitonic form
factors in the sine-Gordon model for arbitrary choice of the parameters.
The regularized multi-soliton form factors could them be used to obtain
correlation functions of direct physical interest, e.g. in condensed
matter physics \cite{Essler:2004}.

The outline of the paper is as follows. In Section 2 the sine-Gordon
model along with its exact form factors are reviewed and  based
on \cite{Lukyanov:1997} integral representations for the form factors
of exponential operators are given. Section 3 is devoted to
the analysis of a certain function which appears in the integral representations.
Giving this function's asymptotic series and identifying its poles
make it possible to analytically continue the integral representations.
In Section 4 explicit formulae are provided for the four-soliton form
factors. Section 5 is devoted to discussion of test results, while 
Section 6 is left for conclusions and outlook.

\section{Form factors in the sine-Gordon model}

\subsection{Definitions, S-matrix and the form factor axioms}

The sine-Gordon model is defined by the classical Lagrangian
\begin{equation}\label{Lagr}
L=\int_{-\infty}^{\infty}dx\left[\frac{1}{2}\partial_{\mu}\varphi\partial^{\mu}\varphi+\frac{m_{0}^{2}}{\beta^{2}}\cos\left(\beta\varphi\right)\right].
\end{equation}
Define the parameter
\begin{equation}\label{xi}
\xi=\frac{\beta^{2}}{1-\beta^{2}}
\end{equation}
which is relevant in the low-energy description of the theory.
The spectrum of the quantum theory contains the soliton-antisoliton
doublet and their bound states, the {}``breathers''. The number of 
breather states ($B_1$, $B_2$, ..., $B_{\bar{N}}$)
is bounded, there are 
$\bar{N}=\left\lfloor \frac{1}{\xi}\right\rfloor $ of them. 
For our purposes it is enough to consider only the solitonic particles
of the spectrum, indexed in the following with $\pm$ (soliton-antisoliton).

\subsubsection{S-matrix}

The S-matrix for the soliton-antisoliton doublet reads
\begin{equation}
\left(\begin{array}{cccc}
S\\
 & S_{T} & S_{R}\\
 & S_{R} & S_{T}\\
 &  &  & S
\end{array}\right),
\end{equation}
with the non-zero elements \cite{Zamolodchikov:1978}
\begin{eqnarray}
S_{++}^{++}\left(\Theta\right) & = & S_{--}^{--}\left(\Theta\right)=S\left(\Theta\right),\\
S_{+-}^{+-}\left(\Theta\right) & = & S_{-+}^{-+}\left(\Theta\right)=S_{T}\left(\Theta\right)=S\left(\Theta\right)\frac{\sinh\frac{\Theta}{\xi}}{\sinh\frac{i\pi-\Theta}{\xi}},\\
S_{+-}^{-+}\left(\Theta\right) & = & S_{-+}^{+-}\left(\Theta\right)=S_{R}\left(\Theta\right)=S\left(\Theta\right)\frac{\sinh\frac{i\pi}{\xi}}{\sinh\frac{i\pi-\Theta}{\xi}},
\end{eqnarray}
where
\begin{multline}
S(\Theta)=-(-1)^N\prod_{k=1}^N\frac{ik\pi\xi+\Theta}{ik\pi\xi-\Theta}\\
\times
\exp\left[-i\int_0^\infty\frac{dt}{t}\sin(\Theta t)\frac{2\sinh\frac{\pi(1-\xi)t}{2}e^{-N\pi\xi t}+(e^{-N\pi\xi t}-1)\left(e^{\frac{\pi(\xi-1)t}{2}}+e^{-\frac{\pi(\xi+1)t}{2}}\right)}{2\sinh\frac{\pi\xi t}{2}\cosh\frac{\pi t}{2}}\right],
\end{multline}
which is independent of the integer $N$, however the integral converges in a larger domain of $\mathbb{C}$ for $N>0$.

\subsubsection{Form factors}

Consider the matrix elements
\begin{equation}
\mathcal{F}_{a_{1}\ldots a_{n}}^{b_{1}\ldots b_{m}}\left(\Theta_{1}',\ldots\Theta_{m}'|\Theta_{1},\ldots\Theta_{n}\right)=\,_{out}\left\langle A^{b_{m}}\left(\Theta_{m}'\right)\ldots A^{b_{1}}\left(\Theta_{1}'\right)|O|A_{a_{1}}\left(\Theta_{1}\right)\ldots A_{a_{n}}\left(\Theta_{n}\right)\right\rangle _{in}
\end{equation}
of the local, hermitian operator $O$ between asymptotic
states. The form factors are defined by
\begin{equation}
\mathcal{F}_{a_{1}\ldots a_{n}}\left(\Theta_{1},\ldots\Theta_{n}\right)=\langle0|O|A_{a_{1}}\left(\Theta_{1}\right)\ldots A_{a_{n}}\left(\Theta_{n}\right)\rangle_{in},
\end{equation}
as the matrix elements of the operator between the vacuum and an $n$-particle
state. Crossing symmetry implies
\begin{equation}
\mathcal{F}_{a_{1}\ldots a_{n}}^{b_{1}\ldots b_{m}}\left(\Theta_{1}',\ldots\Theta_{m}'|\Theta_{1},\ldots\Theta_{n}\right)=\mathcal{F}_{a_{1}\ldots a_{n}(-b_{1})\ldots(-b_{m})}\left(\Theta_{1},\ldots\Theta_{n},\Theta_{1}'+i\pi,\ldots\Theta_{m}'+i\pi\right),
\end{equation}
which is understood as an analytic continuation. The form factors can be reconstructed based on the following axioms.
\begin{enumerate}
\item Analyticity and the physical matrix elements. $\mathcal{F}_{a_{1}\ldots a_{n}}\left(\Theta_{1},\ldots\Theta_{n}\right)$
is analytic in the variables $\Theta_{i}-\Theta_{j}$ inside the physical strip $0<{\rm Im}\,\Theta<2\pi$ except for simple poles. It is the physical matrix element when all $\Theta_{i}$ are real and ordered as $\Theta_{1}<\Theta_{2}<\ldots<\Theta_{n}$. 
\item Relativistic invariance. The form factors satisfy
\begin{equation}
\mathcal{F}_{a_{1}\ldots a_{n}}\left(\Theta_{1}+z,\ldots\Theta_{n}+z\right)=e^{zS\left(O\right)}\mathcal{F}_{a_{1}\ldots a_{n}}\left(\Theta_{1},\ldots\Theta_{n}\right),
\end{equation}
where $S\left(O\right)$ is the spin of the operator $O$.
\item Watson's theorem. The following symmetry properties are satisfied
\begin{equation}
\mathcal{F}_{a_{1}\ldots a_{j}a_{j+1}\ldots a_{n}}\left(\Theta_{1},\ldots\Theta_{j},\Theta_{j+1},\ldots\Theta_{n}\right)=S_{a_{j+1}a_{j}}^{c_{j}c_{j+1}}\left(\Theta_{j+1}-\Theta_{j}\right)\mathcal{F}_{a_{1}\ldots c_{j}c_{j+1}\ldots a_{n}}\left(\Theta_{1},\ldots\Theta_{j+1},\Theta_{j},\ldots\Theta_{n}\right),
\end{equation}
\begin{equation}
\mathcal{F}_{a_{1}\ldots a_{n}}\left(\Theta_{1},\ldots\Theta_{n}+2\pi i\right)=e^{2\pi i\omega\left(O,\Psi\right)}\mathcal{F}_{a_{n}a_{1}\ldots a_{n-1}}\left(\Theta_{n},\Theta_{1},\ldots\Theta_{n-1}\right),
\end{equation}
where the latter is understood as an analytic continuation and $\omega\left(O,\Psi\right)$ is the mutual non-locality index of the operator $O$ and $\Psi$, the {}``elementary'' field, upon which the whole operator product algebra can be constructed.
\item Kinematical poles. $\mathcal{F}_{a_{1}\ldots a_{n}}\left(\Theta_{1},\ldots\Theta_{n}\right)$ has simple poles at $\Theta_{n}=\Theta_{j}+i\pi$ with residues
\begin{alignat}{1}
 -i\mathcal{F}_{a'_{1}\ldots\hat{a}'_{j}\ldots a'_{n}}\left(\Theta_{1},\ldots\hat{\Theta}_{j},\ldots\Theta_{n-1}\right)&\left\{ \delta_{a_{1}}^{a'_{1}}\ldots\delta_{a_{j-1}}^{a'_{j-1}}S_{a_{n-1}c_{1}}^{a'_{n-1}\left(-a_{n}\right)}\left(\Theta_{n-1}-\Theta_{j}\right)S_{a_{n-2}c_{2}}^{a'_{n-2}c_{1}}\left(\Theta_{n-2}-\Theta_{j}\right)\ldots\right.\nonumber \\
 &  \hspace{17em}\times S_{a_{j+1}a_{j}}^{a'_{j+1}c_{n-j-2}}\left(\Theta_{j+1}-\Theta_{j}\right)\nonumber \\
 &  \qquad-e^{2\pi i\omega\left(O,\Psi\right)}S_{c_{1}a_{1}}^{\left(-a_{n}\right)a'_{1}}\left(\Theta_{j}-\Theta_{1}\right)\ldots S_{c_{j-2}a_{j-2}}^{c_{j-3}a'_{j-2}}\left(\Theta_{j}-\Theta_{j-2}\right)\nonumber \\
 & \hspace{12em}\left.\times S_{a_{j}a_{j-1}}^{c_{j-2}a'_{j-1}}\left(\Theta_{j}-\Theta_{j-1}\right)\delta_{a_{j+1}}^{a'_{j+1}}\ldots\delta_{a_{n-1}}^{a'_{n-1}}\right\} .
\end{alignat}
In the absence of bound states these are the only singularities of $\mathcal{F}_{a_{1}\ldots a_{n}}\left(\Theta_{1},\ldots\Theta_{n}\right)$ in the strip $0<{\rm Im}\,\Theta_j<2\pi$ for real $\{\Theta_i\}_{i\neq j}$.

\end{enumerate}

\subsection{Integral representations of multi-soliton form factors}

In \cite{Lukyanov:1997} it is proposed that the $2n$-particle form factors of the exponential operator $e^{ia\varphi}$ in the sine-Gordon model can be represented by
\begin{multline}
\mathcal{F}_{\sigma_1\ldots\sigma_{2n}}(\Theta_1,\ldots,\Theta_{2n})=\langle0|e^{ia\varphi}|A_{\sigma_{2n}}\left(\Theta_{2n}\right)\ldots A_{\sigma_{1}}\left(\Theta_{1}\right)\rangle\\=\mathcal{G}_{a}\langle\langle Z_{\sigma_{2n}}\left(\Theta_{2n}\right)\ldots Z_{\sigma_{1}}\left(\Theta_{1}\right)\rangle\rangle\equiv\mathcal{G}_{a}F_{\sigma_1\ldots\sigma_{2n}}(\Theta_1,\ldots,\Theta_{2n})\label{eq:FF}
\end{multline}
where $\sum_{i=1}^{2n}\sigma_{i}=0$ because of charge conservation and $\mathcal{G}_{a}$ is the vacuum expectation value of the exponential operator \cite{Lukyanov:1997,Lukyanov:1996}. The operators $Z_{\pm}\left(\Theta\right)$ are defined by
\begin{equation}
Z_{+}\left(\Theta\right)=\sqrt{i\frac{\mathcal{C}_{2}}{4\mathcal{C}_{1}}}e^{\frac{a\Theta}{\beta}}e^{i\phi\left(\Theta\right)},
\end{equation}
\begin{multline}
Z_{-}\left(\Theta\right)=\sqrt{i\frac{\mathcal{C}_{2}}{4\mathcal{C}_{1}}}e^{-\frac{a\Theta}{\beta}} \left\{ e^{\frac{i\pi}{2\beta^{2}}}\int_{C_{+}}\frac{d\gamma}{2\pi}e^{\left(1-\frac{2a}{\beta}-\frac{1}{\beta^{2}}\right)\left(\gamma-\Theta\right)}e^{-i\bar{\phi}\left(\gamma\right)}e^{i\phi\left(\Theta\right)}\right. \\ \left.-e^{-\frac{i\pi}{2\beta^{2}}}\int_{C_{-}}\frac{d\gamma}{2\pi}e^{\left(1-\frac{2a}{\beta}-\frac{1}{\beta^{2}}\right)\left(\gamma-\Theta\right)}e^{i\phi\left(\Theta\right)}e^{-i\bar{\phi}\left(\gamma\right)}\right\} .
\end{multline}
Since $\phi(\Theta)$ and $\bar\phi(\gamma)$ are free fields the averaging $\langle\langle\ldots\rangle\rangle$ is performed by the multiplicative Wick's theorem, using
\begin{align}
\langle\langle e^{i\phi\left(\Theta_{2}\right)}e^{i\phi\left(\Theta_{1}\right)}\rangle\rangle&=G\left(\Theta_{1}-\Theta_{2}\right),
\\
\langle\langle e^{i\phi\left(\Theta_{2}\right)}e^{-i\bar{\phi}\left(\Theta_{1}\right)}\rangle\rangle&=W\left(\Theta_{1}-\Theta_{2}\right)=\frac{1}{G\left(\Theta_{1}-\Theta_{2}-\frac{i\pi}{2}\right)G\left(\Theta_{1}-\Theta_{2}+\frac{i\pi}{2}\right)},
\\
\langle\langle e^{-i\bar{\phi}\left(\Theta_{2}\right)}e^{-i\bar{\phi}\left(\Theta_{1}\right)}\rangle\rangle&=\bar{G}\left(\Theta_{1}-\Theta_{2}\right)=\frac{1}{W\left(\Theta_{1}-\Theta_{2}-\frac{i\pi}{2}\right)W\left(\Theta_{1}-\Theta_{2}+\frac{i\pi}{2}\right)}.
\end{align}
The appearing functions and constants are as follows.
\begin{align}
G(\Theta)&=i\mathcal{C}_1\sinh\left(\frac{\Theta}{2}\right)\exp\left\{\int_0^\infty \frac{dt}{t} \frac{\sinh^2t\left(1-\frac{i\Theta}{\pi}\right)\sinh t(\xi-1)}{\sinh(2t)\cosh(t)\sinh(\xi t)}\right\},
\\
\label{Wfun}
W\left(\Theta\right)&=-\frac{2}{\cosh\left(\Theta\right)}\exp\left\{ -2\int_{0}^{\infty}\frac{dt}{t}\frac{\sinh^{2}t\left(1-\frac{i\Theta}{\pi}\right)\sinh t\left(\xi-1\right)}{\sinh2t\sinh\xi t}\right\},
\\
\bar G (\Theta)&=-\frac{\mathcal{C}_2}{4}\xi \sinh\left(\frac{\Theta+i\pi}{\xi}\right)\sinh(\Theta),
\end{align}
\begin{align}
\mathcal{C}_1&=\exp\left\{-\int_0^\infty\frac{dt}{t} \frac{\sinh^2\left(\frac{t}{2}\right)\sinh t(\xi-1)}{\sinh(2t)\cosh(t)\sinh(\xi t)}\right\}=G(-i\pi),
\\
\mathcal{C}_2&=\exp\left\{4\int_0^\infty\frac{dt}{t} \frac{\sinh^2\left(\frac{t}{2}\right)\sinh t(\xi-1)}{\sinh(2t)\sinh(\xi t)}\right\}=\frac{4}{\left[W\left(\frac{i\pi}{2}\right)\xi\sin\left(\frac{\pi}{\xi}\right)\right]^2}.
\end{align}

The integration contours appearing in the consequent expressions for the form
factors are such that the {}``principal poles'' of the $W$-functions are
always between the contour and the real line. (We define the {}``principal
pole'' of $W\left(\Theta\right)$ as the pole located at $\Theta=-\frac{i\pi}{2}$).

For the two-particle form factors it is only necessary to evaluate Eq. (\ref{eq:FF})
for two $Z$ operators. Let $A=-\left(\frac{1}{\xi}+\frac{2a}{\beta}\right),$
then the result is

\begin{alignat}{1}
\langle\langle Z_{+}\left(\Theta_{2}\right) & Z_{-}\left(\Theta_{1}\right)\rangle\rangle=\frac{i\mathcal{C}_{2}}{4\mathcal{C}_{1}}e^{\frac{a}{\beta}\left(\Theta_{2}-\Theta_{1}\right)}G(\Theta_{1}-\Theta_{2})e^{-A\Theta_{1}}\nonumber \\
 & \qquad\times\left\{ e^{\frac{i\pi}{2\beta^{2}}}\int\frac{d\gamma}{2\pi}e^{A\gamma}W\left(\gamma-\Theta_{2}\right)W\left(\Theta_{1}-\gamma\right)-e^{-\frac{i\pi}{2\beta^{2}}}\int\frac{d\gamma}{2\pi}e^{A\gamma}W\left(\gamma-\Theta_{2}\right)W\left(\gamma-\Theta_{1}\right)\right\} ,
\end{alignat}
\begin{alignat}{1}
\langle\langle Z_{-}\left(\Theta_{2}\right) & Z_{+}\left(\Theta_{1}\right)\rangle\rangle=\frac{i\mathcal{C}_{2}}{4\mathcal{C}_{1}}e^{\frac{a}{\beta}\left(\Theta_{1}-\Theta_{2}\right)}G(\Theta_{1}-\Theta_{2})e^{-A\Theta_{2}}\nonumber \\
 & \times\left\{ e^{\frac{i\pi}{2\beta^{2}}}\int\frac{d\gamma}{2\pi}e^{A\gamma}W\left(\Theta_{2}-\gamma\right)W\left(\Theta_{1}-\gamma\right)-e^{-\frac{i\pi}{2\beta^{2}}}\int\frac{d\gamma}{2\pi}e^{A\gamma}W\left(\gamma-\Theta_{2}\right)W\left(\Theta_{1}-\gamma\right)\right\} ,
\end{alignat}

The four-particle form factors can also be obtained through evaluating
Eq. (\ref{eq:FF}) with the result

\begin{eqnarray}
\langle\langle Z_{\sigma_{4}}\left(\Theta_{4}\right)Z_{\sigma_{3}}\left(\Theta_{3}\right)Z_{\sigma_{2}}\left(\Theta_{2}\right)Z_{\sigma_{1}}\left(\Theta_{1}\right)\rangle\rangle & = & \frac{\xi\mathcal{C}_{2}^{3}}{1024\pi^{2}\mathcal{C}_{1}^{2}}G\left(\Theta_{3}-\Theta_{4}\right)G\left(\Theta_{2}-\Theta_{4}\right)G\left(\Theta_{1}-\Theta_{4}\right)\nonumber \\
 &  & \times G\left(\Theta_{2}-\Theta_{3}\right)G\left(\Theta_{1}-\Theta_{3}\right)G\left(\Theta_{1}-\Theta_{2}\right)J_{\sigma_{1}\sigma_{2}\sigma_{3}\sigma_{4}},\label{eq:4sol}
\end{eqnarray}
where

\begin{equation}
J_{\sigma_{1}\sigma_{2}\sigma_{3}\sigma_{4}}=e^{\frac{a}{\beta}\sum_{i=1}^{4}\sigma_{i}\Theta_{i}}e^{-A\sum_{\sigma_{i}=-1}\Theta_{i}}I_{\sigma_{1}\sigma_{2}\sigma_{3}\sigma_{4}}
\end{equation}
and $I_{\sigma_{1}\sigma_{2}\sigma_{3}\sigma_{4}}$'s are given by

\begin{eqnarray}
I_{--++} & = & e^{\frac{i\pi}{\beta^{2}}}\boldsymbol{P}\left(I_{22},I_{31}\right)-\boldsymbol{P}\left(I_{22},I_{40}\right)-\boldsymbol{P}\left(I_{31},I_{31}\right)+e^{-\frac{i\pi}{\beta^{2}}}\boldsymbol{P}\left(I_{31},I_{40}\right),\\
I_{-+-+} & = & e^{\frac{i\pi}{\beta^{2}}}\boldsymbol{P}\left(I_{13},I_{31}\right)-\boldsymbol{P}\left(I_{22},I_{31}\right)-\boldsymbol{P}\left(I_{13},I_{40}\right)+e^{-\frac{i\pi}{\beta^{2}}}\boldsymbol{P}\left(I_{22},I_{40}\right),\\
I_{-++-} & = & e^{\frac{i\pi}{\beta^{2}}}\boldsymbol{P}\left(I_{04},I_{31}\right)-\boldsymbol{P}\left(I_{13},I_{31}\right)-\boldsymbol{P}\left(I_{04},I_{40}\right)+e^{-\frac{i\pi}{\beta^{2}}}\boldsymbol{P}\left(I_{13},I_{40}\right),\\
I_{++--} & = & e^{\frac{i\pi}{\beta^{2}}}\boldsymbol{P}\left(I_{04},I_{13}\right)-\boldsymbol{P}\left(I_{13},I_{13}\right)-\boldsymbol{P}\left(I_{04},I_{22}\right)+e^{-\frac{i\pi}{\beta^{2}}}\boldsymbol{P}\left(I_{13},I_{22}\right),\\
I_{+-+-} & = & e^{\frac{i\pi}{\beta^{2}}}\boldsymbol{P}\left(I_{04},I_{22}\right)-\boldsymbol{P}\left(I_{13},I_{22}\right)-\boldsymbol{P}\left(I_{04},I_{31}\right)+e^{-\frac{i\pi}{\beta^{2}}}\boldsymbol{P}\left(I_{13},I_{31}\right),\\
I_{+--+} & = & e^{\frac{i\pi}{\beta^{2}}}\boldsymbol{P}\left(I_{13},I_{22}\right)-\boldsymbol{P}\left(I_{22},I_{22}\right)-\boldsymbol{P}\left(I_{13},I_{31}\right)+e^{-\frac{i\pi}{\beta^{2}}}\boldsymbol{P}\left(I_{22},I_{31}\right).
\end{eqnarray}
The integrals $I_{ij}$ have four components, $I_{ij,k}$ $k=1,\ldots,4$
and the operation $\boldsymbol{P}$ is defined by
\begin{equation}
\boldsymbol{P}\left(a,b\right)=e^{\frac{i\pi}{\xi}}\left(a_{1}b_{1}-a_{2}b_{2}\right)-e^{-\frac{i\pi}{\xi}}\left(a_{3}b_{3}-a_{4}b_{4}\right).
\end{equation}
$I_{ij,k}$'s read
\begin{equation}
I_{04,k}=\int e^{\left(A+\alpha_{k}\right)x}W\left(\Theta_{4}-x\right)W\left(\Theta_{3}-x\right)W\left(\Theta_{2}-x\right)W\left(\Theta_{1}-x\right)dx,
\end{equation}
\begin{equation}
I_{13,k}=\int e^{\left(A+\alpha_{k}\right)x}W\left(x-\Theta_{4}\right)W\left(\Theta_{3}-x\right)W\left(\Theta_{2}-x\right)W\left(\Theta_{1}-x\right)dx,
\end{equation}

\begin{equation}
I_{22,k}=\int e^{\left(A+\alpha_{k}\right)x}W\left(x-\Theta_{4}\right)W\left(x-\Theta_{3}\right)W\left(\Theta_{2}-x\right)W\left(\Theta_{1}-x\right)dx,
\end{equation}

\begin{equation}
I_{31,k}=\int e^{\left(A+\alpha_{k}\right)x}W\left(x-\Theta_{4}\right)W\left(x-\Theta_{3}\right)W\left(x-\Theta_{2}\right)W\left(\Theta_{1}-x\right)dx,
\end{equation}

\begin{equation}
I_{40,k}=\int e^{\left(A+\alpha_{k}\right)x}W\left(x-\Theta_{4}\right)W\left(x-\Theta_{3}\right)W\left(x-\Theta_{2}\right)W\left(x-\Theta_{1}\right)dx
\end{equation}
with $\alpha_{1}=-1-\frac{1}{\xi}$, $\alpha_{2}=1-\frac{1}{\xi}$,
$\alpha_{3}=-1+\frac{1}{\xi}$, $\alpha_{4}=+1+\frac{1}{\xi}$ coming
from writing $\bar{G}(x)$ as the sum of four exponentials; the contours are as before.

In general, the $2n$-particle form factor is realized as
\begin{alignat}{1}
\langle\langle\prod_{i=1}^{n}Z_{+}\left(\Theta_{i+n}\right)\prod_{i=1}^{n}Z_{-}\left(\Theta_{i}\right)\rangle\rangle= & \left(\frac{i\mathcal{C}_{2}}{8\pi\mathcal{C}_{1}}\right)^{n}e^{\frac{a}{\beta}\sum_{i=1}^{n}\left(\Theta_{i+n}-\Theta_{i}\right)}e^{-A\sum_{i=1}^{n}\Theta_{i}}\prod_{j>i}G\left(\Theta_{i}-\Theta_{j}\right)\nonumber \\
 & \times\int\left\{ \prod_{{\substack{i=1\\ \phantom{j\neq i} }}}^{n}d\gamma_{i}e^{A\gamma_{i}}\left(e^{\frac{i\pi}{2\beta}}W\left(\Theta_{i}-\gamma_{i}\right)-e^{-\frac{i\pi}{2\beta}}W\left(\gamma_{i}-\Theta_{i}\right)\right)\right.\nonumber \\
 & \qquad\qquad\times\left.\prod_{j=1}^{n}W\left(\gamma_{i}-\Theta_{j+n}\right)\prod_{{\substack{j=1\\  j\neq i}}}^n W\left(\text{sign}\left(j-i\right)\left(\gamma_{i}-\Theta_{j}\right)\right)\right\} \cdot\prod_{j>i}\bar{G}\left(\gamma_{j}-\gamma_{i}\right)
\end{alignat}
The last product gives the numerical factor $\left(-\frac{\mathcal{C}_{2}\xi}{16}\right)^{\frac{n(n-1)}{2}}$
and the sum of $4^{\frac{n(n-1)}{2}}$ exponentials containing $\gamma_{i}$'s.
All in all, we have $\left(n+1\right)$ combinations of the $W$-functions,
which must be integrated over with some exponential factors. Note that
the exponential factors do not alter the structure (e.g. the poles)
of the integrands. The other kinds of $2n$-particle form factors
can be obtained e.g. through the symmetry properties of form factors (Watson's
theorem).

The problem with such integrals is that they diverge for either

\begin{equation}
{\rm Re}\, a>\frac{\beta}{2}
\end{equation}
or

\begin{equation}
{\rm Re}\, a<-\frac{1}{\beta}+\frac{\beta}{2}.
\end{equation}
For such choices of $a$ the integrands have essential singularities
at ${\rm Re}\,x\to\pm\infty$. In the next section we prove that $W\left(x\right)$
has an asymptotic series in exponentials of $x$, therefore
the divergent integrals can always be analytically continued to obtain
a finite result. Our strategy is to first deform the integration contours to the real
line, then extract the divergent terms of the integrands in the form of exponentials and give their contributions
exactly by the analytic continuation rules
\begin{eqnarray}
\int_{-\infty}^{0}\exp\left(\alpha x\right)dx & \equiv & +\frac{1}{\alpha},\qquad\alpha\in\mathbb{C},\\
\int_{0}^{+\infty}\exp\left(\alpha x\right)dx & \equiv & -\frac{1}{\alpha},\qquad\alpha\in\mathbb{C}.
\end{eqnarray}
Then if the integral is expected to be analytic and $\int_{0}^{\infty}f\left(x\right)dx$
exists we have
\begin{equation}
\int_{-\infty}^{\infty}f\left(x\right)dx\equiv\sum_{i}\frac{a_{i}}{\alpha_{i}}+\int_{-\infty}^{0}\left(f\left(x\right)-[f]\left(x\right)\right)dx+\int_{0}^{\infty}f(x)dx\label{eq:analc}
\end{equation}
for some $f\left(t\right)$ admitting an asymptotic expansion in exponentials,
\begin{equation}
f\left(x\right)=\sum_{\alpha_{i}<0}a_{i}e^{\alpha_{i}x}+O\left(e^{\alpha_{+}x}\right)\equiv [f](x)+O\left(e^{\alpha_{+}x}\right) ,\qquad x\to-\infty,\quad\alpha_{+}>0.
\end{equation}
The previous equation defines the function $[f](x)$.
The case when $\int_{0}^{\infty}f\left(x\right)dx=\infty$ is similar.

It should be noted that for some combination of the parameters, the analytic continuation
may still produce an infinite result, that is in the case $\alpha_i=0$ for some $i$.  This
happens e.g. for the integral $I_{22,1}$ when $\frac{a}{\beta}=\frac{1}{2}$. These
infinities, however, must and indeed do cancel out from our end results, the form factors, therefore
the $\alpha_{i}=0$ terms in the asymptotic series should be omitted before making the
analytic continuation prescribed in (\ref{eq:analc}).

\section{Analysis of the $W$-function}

\subsection{Asymptotic series}

The function $W\left(x\right)$ is given by (\ref{Wfun}). The asymptotic series of $\cosh(x)^{-1}$ reads as
\begin{equation}
\cosh(x)^{-1}=2e^{-sx}\left(1-e^{-2sx}+e^{-4sx}+\ldots\right),\qquad{\rm Re}\, x\to s\cdot\infty,
\end{equation}
where $\text{s\ensuremath{\equiv}sign}{\rm Re}\, x$ was introduced for convenience.
The exponent of the remaining part of $W(x)$ can be rewritten as
\begin{equation}
\int_{0}^{\infty}\frac{dt}{t}\frac{\sinh\left(\xi-1\right)t}{\sinh\left(\xi t\right)\sinh\left(2t\right)}\left(1-\cosh\left(2t\right)\cos\left(\frac{2tx}{\pi}\right)+i\sinh\left(2t\right)\sin\left(\frac{2tx}{\pi}\right)\right).
\end{equation}
Differentiate the previous formula with respect to $x$ and obtain
\begin{eqnarray}
\frac{2}{\pi}\int_{0}^{\infty}dt\frac{\sinh\left(\xi-1\right)t}{\sinh\left(\xi t\right)\sinh\left(2t\right)}\left(\cosh\left(2t\right)\sin\left(\frac{2tx}{\pi}\right)+i\sinh\left(2t\right)\cos\left(\frac{2tx}{\pi}\right)\right) & =\label{eq:olverhez}\\
\frac{i}{\pi}\left(\int_{-\infty}^\infty \frac{\sinh\left(\xi-1\right)t}{\sinh\left(\xi t\right)}(1-\coth(2t))e^{it\frac{2x}{\pi}}dt\right).
\end{eqnarray}
The asymptotic series of the Fourier integrals was first obtained by
\cite{Olver:1974}, where an analog to Watson's
lemma for Laplace transforms was discussed. Given a function $q\left(t\right)$ with
the asymptotic series near $t=0$:
\begin{equation}
q\left(t\right)=\sum_{n=0}^{\infty}b_{n}t^{n+\lambda-1}
\end{equation}
with some $0<\lambda\leq1$, the asymptotic series of
\begin{equation}
F\left(x\right)=\int_{0}^{\infty}q\left(t\right)e^{\pm itx}dt
\end{equation}
as $x\to\infty$ is given by
\begin{equation}
F\left(x\right)=\sum_{n=0}^{\infty}b_{n}e^{\pm\frac{i\pi}{2}\left(n+\lambda\right)}\Gamma\left(n+\lambda\right)x^{-n-\lambda}+O\left(e^{-\mu x}\right),\qquad x\to\infty,
\end{equation}
where $O\left(e^{-\mu x}\right)$ denotes corrections {}``beyond
all orders'', i.e. exponentially small terms ($\mu>0$). Applying
this construction to the integrals occurring in Eq. (\ref{eq:olverhez})
we get
\begin{equation}
\frac{2}{\pi}\int_{0}^{\infty}dt\frac{\sinh\left(\xi-1\right)t}{\sinh\left(\xi t\right)}\coth\left(2t\right)\sin\left(\frac{2tx}{\pi}\right)=\frac{2}{\pi}\frac{\xi-1}{2\xi}{\rm Im}\,\lim_{\lambda\to0}e^{\frac{i\pi}{2}\lambda}\Gamma\left(\lambda\right)+O\left(e^{-\mu_{1}x}\right)=\frac{\xi-1}{2\xi}+O\left(e^{-\mu_{1}x}\right)
\end{equation}
and
\begin{equation}
\frac{2}{\pi}\int_{0}^{\infty}dt\frac{\sinh\left(\xi-1\right)t}{\sinh\left(\xi t\right)}\cos\left(\frac{2tx}{\pi}\right)=O\left(e^{-\mu_{2}x}\right)
\end{equation}
considering that $q\left(t\right)$ is odd in the first integral and
even in the second which implies that all but the first term of the
first integral disappears because of the ${\rm Im}$/${\rm Re}$ operation,
respectively. For simplicity it was assumed, that $x$ is real. It is easy to
check our statements remain true  if this condition is relaxed.

To find the exponentially small terms the integral is evaluated by
the residue theorem which yields only the residues times $2\pi i$
since the integrands are of small enough order on the half circles
$C_{R}^{\pm}=\left\{ z\,:\,\pm{\rm Im}\, z>0,\,|z|=R\right\} $ as $R\to\infty$
(where $C_{R}^{+}$ is associated with ${\rm Re}\, x>0$ while $C_{R}^{-}$
with ${\rm Re}\, x<0$). The result is
\begin{alignat}{1}
s\left[\frac{\xi-1}{2\xi}+\sum_{k=1}^{\infty}\left(-1\right)^{k+1}\right. & \cot\left(\frac{\pi\xi\left(2k+1\right)}{2}\right)e^{-\left(2k+1\right)sx}+\sum_{k=1}^{\infty}\left(-1\right)^{k}e^{-2ksx}\nonumber \\
 & \left.-\frac{2}{\xi}\sum_{k=1}^{\infty}\sin\left(\frac{\pi k}{\xi}\right)\cot\left(\frac{2\pi k}{\xi}\right)e^{-\frac{2k}{\xi}sx}\right]+\frac{2i}{\xi}\sum_{k=1}^{\infty}\left(-1\right)^{k+1}\sin\left(\frac{\left(\xi-1\right)\pi k}{\xi}\right)e^{-\frac{2k}{\xi}sx}.
\end{alignat}
When integrated with respect to $x$ one gets the exponent of the
$W$-function:

\[
\frac{\xi-1}{2\xi}sx+\sum_{k=1}^{\infty}\frac{1}{k}\sin\left(\frac{\pi k}{\xi}\right)\left[\cot\left(\frac{2\pi k}{\xi}\right)-is\right]e^{-\frac{2k}{\xi}sx}-\sum_{k=1}^{\infty}\frac{1}{k}\left. \begin{cases}
i^{k+1}\cot\left(\frac{\pi\xi k}{2}\right),&\text{ for odd $k$}\\
i^{k},&\text{ for even $k$}
\end{cases}\right\} e^{-ksx}+C_{s}.
\]
The integration constant, $C_{s}$ is determined from the relation

\begin{equation}
\bar{G}(x)=\frac{1}{W\left(x+\frac{i\pi}{2}\right)W\left(x-\frac{i\pi}{2}\right)}
\end{equation}
and the explicit form of the function $\bar{G}(x)$, that is

\begin{equation}
\bar{G}\left(x\right)=-\frac{\mathcal{C}_{2}}{4}\xi\sinh\left(\frac{\Theta+i\pi}{\xi}\right)\sinh\left(\Theta\right),
\end{equation}
with the result

\begin{equation}
-4e^{C_{s}}=\frac{4i}{\sqrt{\mathcal{C}_{2}\xi}}e^{-\frac{is\pi}{2\xi}}.
\end{equation}

Now we are ready to give the asymptotic expansion of $W\left(x\right)$:

\begin{equation}\label{Was}
W\left(x\right)=\frac{4i}{\sqrt{\mathcal{C}_{2}\xi}}e^{-\frac{is\pi}{2\xi}}e^{-\frac{\xi+1}{2\xi}sx}\left(\sum_{l=0}^{\infty}\left(-1\right)^{l}e^{-2lsx}\right)\prod_{k=1}^{\infty}\left(\left[\sum_{l=0}^{\infty}\frac{a_{k}^{l}}{l!}e^{-\frac{2kl}{\xi}sx}\right]\left[\sum_{l=0}^{\infty}\frac{b_{k}^{l}}{l!}e^{-klsx}\right]\right),\quad {\rm Re}\,x\to s\cdot\infty,
\end{equation}
where the coefficients depend only on $\xi$ and $\text{s\ensuremath{\equiv}sign}\left({\rm Re}\,x\right)$
and are expressed as

\begin{equation}
a_{k}=\frac{1}{k}\left[\frac{\cos\left(\frac{2\pi k}{\xi}\right)}{2\cos\left(\frac{\pi k}{\xi}\right)}-is\sin\left(\frac{\pi k}{\xi}\right)\right]
\end{equation}
and
\begin{equation}
b_{k}=-\frac{1}{k} \begin{cases}
i^{k+1}\cot\left(\frac{\pi\xi k}{2}\right)&\text{ for odd $k$,}\\
i^{k}&\text{ for even $k$.}
\end{cases}
\end{equation}

Note, that, strictly speaking, our expansion is limited to the case
of irrational $\xi$ parameters since otherwise the coefficients $a_{k}$
and $b_{k}$ always become infinite for some $k$. However, such infinities
can be shown to cancel out.

Let $\xi=\frac{n_{1}}{n_{2}}$, where $n_{1}$ and $n_{2}$ are relative
primes. We have singular $a_{k}$'s whenever $\frac{2kn_{2}}{n_{1}}=\text{odd}$,
which immediately implies $n_{1}=\text{even}$ and $n_{2}=\text{odd}$
and the $N$th singular $a$-coefficient is indexed by $\frac{Nn_{1}}{2}$,
where $N$ is necessarily odd. On the other hand $b_{l}$ is singular
if $\frac{ln_{1}}{n_{2}}=\text{even}$ while $l$ is odd, implying
again $n_{1}=\text{even}$ and $n_{2}=\text{odd}$
and the $N$th singular $b$-coefficient is indexed by $Nn_{2}$.
Now in both cases the $N$th singular term contribute terms of order
$e^{-Nn_{2}x}$ to the exponent of $W\left(x\right)$. All that remains
is to show that the $N$th diverging coefficients cancel each other.
To see this let $\xi=\frac{n_{1}}{n_{2}}\left(1+\varepsilon\right)$
or equivalently

\begin{eqnarray}
n_{1} & \to & n_{1}\left(1+\varepsilon\right),\\
n_{2} & \to & n_{2}\left(1-\varepsilon\right),
\end{eqnarray}
resulting in
\begin{eqnarray}
{\rm Re}\, a_{Nn_{1}/2}=\frac{1}{Nn_{1}}\frac{\cos\left(\pi Nn_{2}\right)}{\cos\left(\pi Nn_{2}/2\right)} & \longrightarrow & -\frac{1}{Nn_{1}}\frac{1}{\cos\left(\pi Nn_{2}\left(1-\varepsilon\right)/2\right)},\\
b_{Nn_{2}}=-\frac{1}{Nn_{2}}i^{Nn_{2}+1}\frac{\cos\left(\pi Nn_{1}/2\right)}{\sin\left(\pi Nn_{1}/2\right)} & \longrightarrow & -\frac{1}{Nn_{2}}\frac{\left(-1\right)^{\frac{N\left(n_{1}+n_{2}\right)+1}{2}}}{\sin\left(\pi Nn_{1}\left(1+\varepsilon\right)/2\right)}
\end{eqnarray}
which if expanded in $\varepsilon$ yield
\begin{equation}
{\rm Re}\, a_{Nn_{1}/2}=b_{Nn_{2}}+O\left(\varepsilon\right)=\left(-1\right)^{\frac{Nn_{2}+3}{2}}\frac{4}{\pi N^{2}n_{1}n_{2}}\frac{1}{\varepsilon}+O\left(\varepsilon\right),
\end{equation}
which agrees for $\varepsilon\to0$. Because of the cancellation the
following rules can be formulated for rational $\xi$'s:

\begin{equation}
a_{k}=-\frac{is}{k}\text{ for }\frac{2k}{\xi}=\text{odd},
\end{equation}
\begin{equation}
b_{k}=0\text{ for }k=\text{odd and }k\xi=\text{even}.
\end{equation}

\subsection{Poles}

With the asymptotic expansion at hand the divergences of the integral
representations can be readily remedied. However there is another
issue with the integrals containing $W$-functions. In fact, $W(x)$
has a number of poles on the line ${\rm Re}\, x=0$. When the integration
contour is fixed (which is the desired scenario), poles can cross
it and one needs to analytically continue the result by adding the residue
contributions of the crossing poles. In the followings we determine
the poles of $W(x)$. The poles of $W(x)$ are easily extracted from
the identity (\cite{Takacspc}, but also follows from a similar representation
of $G(x)$, given in \cite{FeherTakacs})
\begin{alignat}{2}\label{Wreg}
W\left(x\right)=-\frac{2}{\cosh x} & \prod_{k=1}^{N}\frac{\Gamma\left(1+\frac{2k-\frac{5}{2}+\frac{ix}{\pi}}{\xi}\right)\Gamma\left(1+\frac{2k-\frac{1}{2}-\frac{ix}{\pi}}{\xi}\right)\Gamma\left(\frac{2k-\frac{1}{2}}{\xi}\right)^{2}}{\Gamma\left(1+\frac{2k-\frac{3}{2}}{\xi}\right)^{2}\Gamma\left(\frac{2k+\frac{1}{2}-\frac{ix}{\pi}}{\xi}\right)\Gamma\left(\frac{2k-\frac{3}{2}+\frac{ix}{\pi}}{\xi}\right)}\nonumber \\
 & \hspace{3cm}\times\exp\left\{ -2\int_{0}^{\infty}\frac{dt}{t}\frac{e^{-4Nt}\sinh^{2}t\left(1-\frac{i x}{\pi}\right)\sinh t\left(\xi-1\right)}{\sinh2t\sinh\xi t}\right\} .
\end{alignat}
They originate from the poles of the gamma functions and the roots
of $\cosh x$. It is apparent that ${\rm Re}\, x=0$ for every pole. We are
interested in the poles of $W\left(x-x_{0}\right)$ which cross the
real line (or the original integration contour) when $\xi$ is decreased.
The previous equation yields two infinite series of poles given by
\begin{eqnarray}
\xi+2k-\frac{5}{2}+\frac{ix_{1,k,n}}{\pi} & = & -n\xi,\label{xipeq}\\
\xi+2k-\frac{1}{2}-\frac{ix_{2,k,n}}{\pi} & = & -n\xi,
\end{eqnarray}
$n$ being a non-negative integer and $k$ being a positive number.
We have the following estimates for the series of poles:
\begin{eqnarray}
{\rm Im}\, x_{1,k,n} & > & \frac{\left(4k-5\right)\pi}{2},\\
{\rm Im}\, x_{2,k,n} & < & \frac{\left(1-4k\right)\pi}{2}.
\end{eqnarray}
For $|{\rm Im}\, x_{0}|\leq\frac{3}{2}\pi$ the only poles that can cross
the real line are
\begin{equation}
x_{n}\equiv x_{1,1,n}=i\pi\left(n\xi-\frac{1}{2}\right),\qquad n=1,2,\ldots.\label{eq:xipoles}
\end{equation} 
and no poles can
cross the original contour, which intersects the ${\rm Im}\, x=0$ line at
$x-x_{0}=-\frac{i\pi}{2}-\varepsilon$, $\varepsilon\to0$.  
With reference to the form factors, note that because of Watson's
theorem it is enough to give a calculation method when all
the rapidities satisfy $|{\rm Im}\,\Theta_{i}|\leq\pi$. Thus it is not necessary
to analyze further the $\xi$-dependent poles of $W\left(x-x_{0}\right)$,
that is covering the case $|{\rm Im}\, x_{0}|>\pi$.

\section{Implementation of the four-soliton form factor formula}

As an example of the machinery outlined in Section 2 in this
section the implementation of the four-soliton form factors is discussed.
Implementing Eq. (\ref{eq:4sol}) is non-trivial only in the
calculation of the integrals $I_{ij,k}$. 

First, by Cauchy's theorem
we deform the integration contour to the real line.
For this we need to identify the poles between the real line and the
original contour, which consists of the the principal poles and (possibly) several
$\xi$-dependent poles (given by Eq. (\ref{eq:xipoles})) of $W$-functions.
The principal poles give the following contributions to $I_{22,k}$:

\begin{eqnarray}
P_{1,k} & = & -\frac{4}{\pi\sqrt{\mathcal{C}_{2}}}e^{\left(A+\alpha_{k}\right)\left(\Theta_{1}+\frac{i\pi}{2}\right)}W\left(\Theta_{1}-\Theta_{4}+\frac{i\pi}{2}\right)W\left(\Theta_{1}-\Theta_{3}+\frac{i\pi}{2}\right)W\left(\Theta_{2}-\Theta_{1}-\frac{i\pi}{2}\right),\\
P_{2,k} & = & -\frac{4}{\pi\sqrt{\mathcal{C}_{2}}}e^{\left(A+\alpha_{k}\right)\left(\Theta_{2}+\frac{i\pi}{2}\right)}W\left(\Theta_{2}-\Theta_{4}+\frac{i\pi}{2}\right)W\left(\Theta_{2}-\Theta_{3}+\frac{i\pi}{2}\right)W\left(\Theta_{1}-\Theta_{2}-\frac{i\pi}{2}\right),\\
P_{3,k} & = & -\frac{4}{\pi\sqrt{\mathcal{C}_{2}}}e^{\left(A+\alpha_{k}\right)\left(\Theta_{3}-\frac{i\pi}{2}\right)}W\left(\Theta_{3}-\Theta_{4}-\frac{i\pi}{2}\right)W\left(\Theta_{2}-\Theta_{3}+\frac{i\pi}{2}\right)W\left(\Theta_{1}-\Theta_{3}+\frac{i\pi}{2}\right),\\
P_{4,k} & = & -\frac{4}{\pi\sqrt{\mathcal{C}_{2}}}e^{\left(A+\alpha_{k}\right)\left(\Theta_{4}-\frac{i\pi}{2}\right)}W\left(\Theta_{4}-\Theta_{3}-\frac{i\pi}{2}\right)W\left(\Theta_{2}-\Theta_{4}+\frac{i\pi}{2}\right)W\left(\Theta_{1}-\Theta_{4}+\frac{i\pi}{2}\right),
\end{eqnarray}
if ${\rm Im}\,\Theta_{1}>-\frac{\pi}{2}$, ${\rm Im}\,\Theta_{2}>-\frac{\pi}{2}$,
${\rm Im}\,\Theta_{3}<+\frac{\pi}{2}$, ${\rm Im}\,\Theta_{4}<+\frac{\pi}{2}$,
respectively. The $\xi$-dependent poles yield

\begin{eqnarray}
X_{1,k} & = & \sum_{n=1}^{N_{1}}2\pi ir_{n}e^{\left(A+\alpha_{k}\right)\left(\Theta_{1}-x_{n}\right)}W\left(\Theta_{1}-\Theta_{4}-x_{n}\right)W\left(\Theta_{1}-\Theta_{3}-x_{n}\right)W\left(\Theta_{2}-\Theta_{1}+x_{n}\right)\\
X_{2,k} & = & \sum_{n=1}^{N_{2}}2\pi ir_{n}e^{\left(A+\alpha_{k}\right)\left(\Theta_{2}-x_{n}\right)}W\left(\Theta_{2}-\Theta_{4}-x_{n}\right)W\left(\Theta_{2}-\Theta_{3}-x_{n}\right)W\left(\Theta_{1}-\Theta_{2}+x_{n}\right)\\
X_{3,k} & = & \sum_{n=1}^{N_{3}}2\pi ir_{n}e^{\left(A+\alpha_{k}\right)\left(\Theta_{3}+x_{n}\right)}W\left(\Theta_{3}-\Theta_{4}+x_{n}\right)W\left(\Theta_{2}-\Theta_{3}-x_{n}\right)W\left(\Theta_{1}-\Theta_{3}-x_{n}\right)\\
X_{4,k} & = & \sum_{n=1}^{N_{4}}2\pi ir_{n}e^{\left(A+\alpha_{k}\right)\left(\Theta_{4}+x_{n}\right)}W\left(\Theta_{4}-\Theta_{3}+x_{n}\right)W\left(\Theta_{2}-\Theta_{4}-x_{n}\right)W\left(\Theta_{1}-\Theta_{4}-x_{n}\right)
\end{eqnarray}
where $N_{1,2}=\left\lfloor \left({\rm Im}\,\Theta_{1,2}/\pi+1/2\right)/\xi\right\rfloor $
and $N_{3,4}=\left\lfloor \left(-{\rm Im}\,\Theta_{3,4}/\pi+1/2\right)/\xi\right\rfloor $,
$x_{n}$ are defined by Eq. (\ref{eq:xipoles}) and $r_{n}$ is the
residue of $W\left(x\right)$ at $x_{n}$, calculated numerically
by the definition:

\begin{equation}\label{resid}
r_{n}=\lim_{x\to x_{n}}\left(x-x_{n}\right)W\left(x\right).
\end{equation}

Second, the integrals, $I_{ij,k}$ with the deformed contours are evaluated by the analytic continuation
formula (\ref{eq:analc}). In conclusion one gets
\begin{equation}
I_{22,k}=\sum_i\frac{a_i}{\alpha_i} +\int_{-\infty}^{0}\left(A\left(x\right)-[A](x)\right)dx+\int_{0}^{\infty}A(x)dx+\sum_{n=1}^{4}\left(P_{n,k}+X_{n,k}\right),\label{eq:I11k}
\end{equation}
with
\begin{align}
A\left(x\right)&=\exp\left[\left(A+\alpha_{k}\right)x\right]W\left(x-\Theta_{4}\right)W\left(x-\Theta_{3}\right)W\left(\Theta_{2}-x\right)W\left(\Theta_{1}-x\right)\\
&=[A](x)+O(e^{\alpha_+ x})=\sum_{\alpha_i<0} a_i e^{\alpha_ix}+O(e^{\alpha_+ x}),\qquad {\rm Re}\,x\to-\infty.
\end{align}

Upon generalization to the $2n$-soliton form factors the only non-trivial
component of this procedure is the determination of the residues picked
up when deforming the contour. For an integrand
\begin{equation}
A\left(x\right)=e^{Bx}\prod_{i=1}^{N}W\left(s_{i}\left(x-\Theta_{i}\right)\right),\qquad s_i=\pm1
\end{equation}
we have
\begin{equation}
\int_{C}A(x)dx=\int_{-\infty}^\infty A(x)dx+P+X,
\end{equation}
with the residue contributions
\begin{equation}
P=-\frac{4}{\pi\sqrt{\mathcal{C}_{2}}}\sum_{i=1}^{N}\Theta\left[-s_{i}{\rm Im}\,\Theta_{i}+\frac{\pi}{2}\right]e^{B\left(\Theta_{i}-s_{i}\frac{i\pi}{2}\right)}\prod_{j\neq i}W\left(s_{j}\left[\Theta_{i}-\Theta_{j}-s_{i}\frac{i\pi}{2}\right]\right),
\end{equation}
and
\begin{equation}
X=\sum_{i=1}^{N}\sum_{n=1}^{N_{i}}2\pi ir_{n}e^{B\left(\Theta_{i}+s_{i}r_{n}\right)}\prod_{j\neq i}W\left(s_{j}\left[\Theta_{i}-\Theta_{j}+s_{i}r_{n}\right]\right),\qquad N_{i}=\left\lfloor \frac{\pi-2s_{i}{\rm Im}\,\Theta_{i}}{2\pi\xi}\right\rfloor .
\end{equation}

To conclude this section, we give some details of the Mathematica \cite{Mathematica} package {\tt SGFF.M}.
After the above, only one element remains that is not straightforward in the implementation: the calculation
of the asymptotic series of the product of several $W$-functions.
Simple products of the series quickly produce an intractable number
of terms, most of which are inaccurate (higher order terms, to which further
orders in the series of the constituent functions would contribute). Our solution makes
use of 2-by-$n$ matrices containing the coefficients and the exponents
of the terms in the asymptotic series. We have the following key procedures in the package {\tt SGFF.M}.
\begin{itemize}
\item $\mathtt{PRO}\left[\mathfrak{A},\mathfrak{B}\right]$ calculates the 2-by-$m$ matrix corresponding to the product of asymptotic series $\mathfrak{A}$, $\mathfrak{B}$ including orders only with accurate coefficients.

\item $\mathtt{SHI}\left[\mathfrak{A},a\right]$ generates the 2-by-$n$ matrix corresponding to the asymptotic series of $f\left(x+a\right)$ from that of $f\left(x\right)$ (i.e. $\mathfrak{A}$).

\item $\mathtt{AInt}\left[\mathfrak{A}\right]$ gives the integral of the asymptotic series corresponding to $\mathfrak{A}$ on the negative half-line.

\item $\mathtt{Asyfun}\left[\mathfrak{A},x\right]$ yields the value of the asymptotic series corresponding to $\mathfrak{A}$ at $x$.

\end{itemize}
In terms of these procedures $I_{22,k}$ is calculated as
\begin{equation}
I_{22,k}=\mathtt{AInt}\left[\mathfrak{A}\right]+\int_{-\infty}^{0}\left(A\left(x\right)-\mathtt{Asyfun}\left[\mathfrak{A},x\right]\right)dx+\int_{0}^{\infty}A(x)dx+\sum_{n=1}^{4}\left(P_{n,k}+X_{n,k}\right),\label{eq:I11k}
\end{equation}
with $A(x)$ as before, and
\begin{equation}
\mathfrak{A}=\mathtt{PRO}\left[\mathtt{PRO}\left(\mathtt{PRO}\left[\mathtt{SHI}\left[\mathfrak{W}^{*},-\Theta_{4}\right],\mathtt{SHI}\left[\mathfrak{W}^{*},-\Theta_{3}\right]\right],\mathtt{SHI}\left(\mathfrak{W},-\Theta_{2}\right]\right],\mathtt{SHI}\left[\mathfrak{W},-\Theta_{1}\right]\right],
\end{equation}
$\mathfrak{W}$ being the asymptotic series of $W(x)$ for ${\rm Re}\,x\geq0$. The remaining integrals are performed by the routine $\mathtt{NIntegrate}$.

From a practical point of view one should note, that the second term in (\ref{eq:I11k}) can be numerically unstable. On the other hand, provided that the truncated asymptotic series is a good enough approximation of $A(x)$ for $x<0$ the contribution of this term can be neglected altogether. Therefore, we omitted this term from our code and supposed that the input rapidities are big enough for this to cause no harm. This can be assumed safely, since Lorentz invariance implies that the rapidities can be shifted by an arbitrary real number.

Considering now the general case of $2n$-particle form factors, we give the asymptotic
series of 
\begin{equation}\label{integrand}
A\left(x\right)=e^{Bx}\prod_{i=1}^{N}W\left(s_{i}\left(x-\Theta_{i}\right)\right),
\end{equation}
diverging for $x\to-\infty$, as
\[
\mathfrak{A}=\mathtt{PRO}_{i=1}^{N}\left[\mathtt{SHI}\left[\mathtt{Co}_{s_{i}}\left(\mathfrak{W}\right),-\Theta_{i}\right]\right],\qquad\mathtt{Co}_{s_{i}}\left(\mathfrak{W}\right)=\left\{ \begin{array}{c}
\mathfrak{W}\phantom{^{^{*}}},\quad s_{i}=+1\\
\mathfrak{W}^{*},\quad s_{i}=-1
\end{array}\right.
\]

The main functions available in {\tt SGFF.M} are to calculate the two-, four- and six-soliton form factors.

\begin{itemize}
\item $\mathtt{FF2}[\Theta_1,\Theta_2]$ gives the two-particle form factors
$$
\{F_{+-}(\Theta_{1,i}-\Theta_{2,i}),F_{-+}(\Theta_{1,i}-\Theta_{2,i})\}
$$
where $\Theta_{1}$ and $\Theta_{2}$ are arrays of the same length with elements $\Theta_{a,i}$ ($a=1,2$, $i=1,2,\ldots N$) rapidities where the two-soliton form factors are to be evaluated.

\item $\mathtt{FF4}[\Theta_1,\Theta_2,\Theta_3,\Theta_4]$ gives the four-particle form factors
$$
\{F_{--++}(\Theta_{1,i},\Theta_{2,i},\Theta_{3,i},\Theta_{4,i}),F_{-+-+}(\ldots),F_{+--+}(\ldots),F_{-++-}(\ldots),F_{+-+-}(\ldots),F_{++--}(\ldots)\}.
$$

\item $\mathtt{FF6}[\Theta_1,\Theta_2,\Theta_3,\Theta_4,\Theta_5,\Theta_6]$ gives the six-particle form factors
\begin{align*}
\{F_{+++---}&(\Theta_{1,i},\Theta_{2,i},\Theta_{3,i},\Theta_{4,i},\Theta_{5,i},\Theta_{6,i}),\,
F_{++-+--}(\ldots),\,F_{++--+-}(\ldots),\,F_{++---+}(\ldots),\\
&F_{+-+-+-}(\ldots),\,F_{+-++--}(\ldots),\,F_{+-+--+}(\ldots),\,F_{+--++-}(\ldots),\,F_{+--+-+}(\ldots),\\
&F_{+---++}(\ldots),\,F_{-+++--}(\ldots),\,F_{-++-+-}(\ldots),\,F_{-++--+}(\ldots),\,F_{-+-++-}(\ldots),\\
&F_{-+--++}(\ldots),\,F_{-+-+-+}(\ldots),\,F_{--+++-}(\ldots),\,F_{--++-+}(\ldots),\,F_{--+-++}(\ldots),\,F_{---+++}(\ldots)
\}.
\end{align*}

\item $\mathtt{FF2p}[\Theta_1,\Theta_2]$ gives the two-particle form factors for physical rapidities, i.e. ones with imaginary parts of $\pm \pi$.

\item $\mathtt{FF4p}[\Theta_1,\Theta_2,\Theta_3,\Theta_4]$ gives the four-particle form factors for physical rapidities.

\item $\mathtt{FF6p}[\Theta_1,\Theta_2,\Theta_3,\Theta_4,\Theta_5,\Theta_6]$ gives the six-particle form factors for physical rapidities.

\end{itemize}

Note that accurate results can only be expected when all rapidities have big enough positive real parts and imaginary part in the interval $[-\pi,\pi]$. The functions calculating form factors only at physical rapidities are considerably faster compared to the general ones if the form factors are needed in more than one points; they calculate the necessary $W$-function values for the integrals only once as part of the initialization.

The parameters that can be specified in $\tt SGFF.M$ are the following, which can be edited e.g. in Mathematica before loading the package.

\begin{itemize}

\item $\mathtt{\xi}$ is the IR parameter (\ref{xi}).

\item $\mathtt{aover\beta}$ is the ratio of the parameter $a$ appearing in the operator $O=e^{i a\varphi}$ and the UV parameter $\beta$ of the Lagrangian.

\item $\mathtt{NN}$ is the regularization parameter for the $G-$ and $W-$functions denoted by $N$ in the formula (\ref{Wreg}).

\item $\mathtt{Na}$ is the maximum number of terms treated in the individual asymptotic series in the formula (\ref{Was}).

\item $\mathtt{Ni}$ is the number of interpolation points used to calculate the integrands of type (\ref{integrand}). When evaluating the form factors at general rapidities, mainly $\mathtt{Ni}$ determines the time of evaluation. However, it is this parameter that influences the accuracy the most, as well. A safe choice is $\mathtt{Ni}=2000$.

\item $\mathtt{\varepsilon}$ is a technical parameter for the calculation of the residues (\ref{resid}), $\varepsilon=x-x_n$.

\item $\mathtt{aa}$ and $\mathtt{bb}$ are the lower and upper bounds of the integrals of the type $\int_0^\infty A(x)dx$ in (\ref{eq:I11k})

\end{itemize}

\section{Tests}

The four-particle form factor $F_{--++}$ was checked against the free fermion point result (omitting the vacuum expectation value)
\begin{alignat}{1}
\langle\langle Z_{+}\left(\Theta_{4}\right)Z_{+}\left(\Theta_{3}\right)Z_{-}\left(\Theta_{2}\right)Z_{-}\left(\Theta_{1}\right)\rangle\rangle= & \sin^{2}\left(\sqrt{2}\pi a\right)e^{\sqrt{2}a\left(\Theta_{4}+\Theta_{3}-\Theta_{2}-\Theta_{1}\right)}\\
 & \times\frac{\sinh\left(\frac{\Theta_{1}-\Theta_{2}}{2}\right)\sinh\left(\frac{\Theta_{3}-\Theta_{4}}{2}\right)}{\cosh\left(\frac{\Theta_{3}-\Theta_{1}}{2}\right)\cosh\left(\frac{\Theta_{3}-\Theta_{2}}{2}\right)\cosh\left(\frac{\Theta_{4}-\Theta_{1}}{2}\right)\cosh\left(\frac{\Theta_{4}-\Theta_{2}}{2}\right)}.\nonumber 
\end{alignat}
We do not show test results for this formula since our calculations agreed with the exact results to the machine precision (of 15 digits).

Also, we investigated whether the numerically obtained form factors satisfy the form factor axioms. In the four-particle case the equation

\begin{equation}\label{W4_0}
F_{\sigma_{1}\sigma_{2}\sigma_{3}\sigma_{4}}\left(\Theta_{1}+z,\Theta_{2}+z,\Theta_{3}+z,\Theta_{4}+z\right)=F_{\sigma_{1}\sigma_{2}\sigma_{3}\sigma_{4}}\left(\Theta_{1},\Theta_{2},\Theta_{3},\Theta_{4}\right)
\end{equation}
must hold, which was checked. Watson's theorem is another axiom, e.g. in the form

\begin{equation}\label{W4_1}
F_{--++}\left(\Theta_{1},\Theta_{2},\Theta_{3},\Theta_{4}\right)=S_{+-}^{-+}\left(\Theta_{3}-\Theta_{2}\right)F_{--++}\left(\Theta_{1},\Theta_{3},\Theta_{2},\Theta_{4}\right)+S_{+-}^{+-}\left(\Theta_{3}-\Theta_{2}\right)F_{-+-+}\left(\Theta_{1},\Theta_{3},\Theta_{2},\Theta_{4}\right),
\end{equation}
and
\begin{equation}\label{W4_2}
F_{--++}\left(\Theta_{1},\Theta_{2},\Theta_{3},\Theta_{4}+2\pi i\right)=e^{2\pi i\omega}F_{+--+}\left(\Theta_{4},\Theta_{1},\Theta_{2},\Theta_{3}\right),
\end{equation}
$\omega=\frac{a}{\beta}$ being the mutual non-locality index.

The residues of the kinematic poles of the four-particle form factors were also checked by:
\begin{equation}\label{kin4}
i\lim_{\Theta_{4}\to\Theta_{2}+i\pi}\left(\Theta_{4}-\Theta_{2}-i\pi\right)F_{--++}\left(\Theta_{1},\Theta_{2},\Theta_{3},\Theta_{4}\right)=F_{-+}\left(\Theta_{1}-\Theta_{3}\right)\left[S_{+-}^{+-}\left(\Theta_{3}-\Theta_{2}\right)-e^{2\pi i\omega}S_{--}^{--}\left(\Theta_{2}-\Theta_{1}\right)\right].
\end{equation}
Testing the kinematical poles is especially important: for
the cases when the two-soliton form factor is known explicitly (e.g. for
half-integer $\frac{a}{\beta}$), equation (\ref{kin4}) gives the only check
that is independent of numerical integrals and their analytic continuations. E.g. for $\frac{a}{\beta}=1$
the two-particle form factors are known to be \cite{Lukyanov:1997}
\begin{equation}
F_{\mp\pm}^\beta(\Theta)=\frac{G(\Theta)}{G(-i\pi)}\cot\left(\frac{\pi\xi}{2}\right)\frac{4i\cosh\left(\frac{\Theta}{2}\right)e^{\mp\frac{\Theta+i\pi}{2\xi}}}{\xi\sinh\left(\frac{\Theta+i\pi}{\xi}\right)}.
\end{equation}

In Tables 1 and 2 we listed test results for the four-particle form factors. One can see that magnitude of the error varies greatly for different scenarios. This is because we work in fixed precision (double precision) and the integrals appearing in the formulas can assume values of very different magnitudes and rounding errors can get magnified.

\begin{table}
\centering
\begin{tabular}{|c|c|c|}
  \hline
& LHS & RHS\\
\hline
  (\ref{W4_0}), $\xi=2.23$ &$0.45330-1.4092i$ &$0.45336-1.4093i$\\
  (\ref{W4_0}), $\xi=0.34$ &$0.00089-0.051i$ &$0.00091-0.049i$\\
  (\ref{W4_1}), $\xi=2.23$ &$0.453360-1.4093198i$ &$0.453358-1.4093196i$\\
  (\ref{W4_1}), $\xi=0.34$ &$0.0009063-0.04937438i$ &$0.0009065-0.04937441i$\\
  (\ref{W4_2}), $\xi=2.23$ &$-0.04255089122137+0.03246926430660i$ &$-0.04255089122139+0.03246926430663i$ \\
  (\ref{W4_2}), $\xi=0.34$ &$-0.043292833089+0.00219194033i$ &$-0.043292833083+0.00219194037i$ \\
  \hline
\end{tabular}\label{tab4}
\caption{Comparison of the LHS's and RHS's of the form factor axioms (\ref{W4_0}) (where $z=1$ was taken), (\ref{W4_1}), (\ref{W4_2}) in the four-soliton case.
The rapidities were chosen to be  $\Theta_{1}=7.6,\,\Theta_{2}=7,\,\Theta_{3}=7.2,$ and $\Theta_{4}=6-i\pi$. In all the tests $\frac{a}{\beta}=\frac{5}{4}$ was set.}
\end{table}

\begin{table}
\centering
\begin{tabular}{|c|c|c|}
  \hline
& LHS & RHS\\
\hline
  (\ref{kin4}), $\xi=2.23$ &$\phantom{-}0.8211182+0.7147548i$ & $\phantom{-}0.8211175+0.7147545i$\\
  (\ref{kin4}), $\xi=1.17$ &$-0.2812726+0.0213804i$ &$-0.2812724+0.0213801i$ \\
  (\ref{kin4}), $\xi=0.34$ &$-0.4726029- 0.6620907i$ &$-0.4726070-  0.6620917i$ \\
  \hline
\end{tabular}\label{tab41}
\caption{Comparison of residues of four-particle form factors with exact results (\ref{kin4}). We took  $\frac{a}{\beta}=1$ and for the rapidities
 $\Theta_{1}=7.6,\,\Theta_{2}=7,\,\Theta_{3}=7.2,$ and $\Theta_4=7+10^{-8}+i\pi$.}
\end{table}

We also implemented the 6-particle form factors. The numerical results
agreed with the exact results for the free fermion point, where the
6-particle form factor $F_{---+++}$ reads
\begin{alignat}{1}
\frac{-i\sin^{3}\left(\sqrt{2}\pi a\right)e^{\sqrt{2}a\left(\Theta_{6}+\Theta_{5}+\Theta_{4}-\Theta_{3}-\Theta_{2}-\Theta_{1}\right)}\sinh\left(\frac{\Theta_{1}-\Theta_{2}}{2}\right)\sinh\left(\frac{\Theta_{4}-\Theta_{5}}{2}\right)}{\cosh\left(\frac{\Theta_{4}-\Theta_{1}}{2}\right)\cosh\left(\frac{\Theta_{4}-\Theta_{2}}{2}\right)\cosh\left(\frac{\Theta_{4}-\Theta_{3}}{2}\right)\cosh\left(\frac{\Theta_{5}-\Theta_{1}}{2}\right)\cosh\left(\frac{\Theta_{5}-\Theta_{2}}{2}\right)}\nonumber \\
\times\frac{\sinh\left(\frac{\Theta_{1}-\Theta_{3}}{2}\right)\sinh\left(\frac{\Theta_{4}-\Theta_{6}}{2}\right)\sinh\left(\frac{\Theta_{2}-\Theta_{3}}{2}\right)\sinh\left(\frac{\Theta_{5}-\Theta_{6}}{2}\right)}{\cosh\left(\frac{\Theta_{5}-\Theta_{3}}{2}\right)\cosh\left(\frac{\Theta_{6}-\Theta_{1}}{2}\right)\cosh\left(\frac{\Theta_{6}-\Theta_{2}}{2}\right)\cosh\left(\frac{\Theta_{6}-\Theta_{3}}{2}\right)}
\end{alignat}
In addition, the following kinematic pole equation was tested:
\begin{alignat}{1}
i\lim_{\Theta_{6}\to\Theta_{3}+i\pi} & \left(\Theta_{6}-\Theta_{3}-i\pi\right)F_{---+++}\left(\Theta_{1},\Theta_{2},\Theta_{3},\Theta_{4},\Theta_{5},\Theta_{6}\right)=\nonumber \\
 & F_{--++}\left(\Theta_{1},\Theta_{2},\Theta_{4},\Theta_{5}\right)\left[S_{+-}^{+-}\left(\Theta_{5}-\Theta_{3}\right)S_{+-}^{+-}\left(\Theta_{4}-\Theta_{3}\right)-e^{2\pi i\omega}S_{--}^{--}\left(\Theta_{3}-\Theta_{1}\right)S_{--}^{--}\left(\Theta_{3}-\Theta_{2}\right)\right].\label{kin6}
\end{alignat}
Watson's theorem requires e.g.
\begin{eqnarray}
F_{---+++}\left(\Theta_{1},\Theta_{2},\Theta_{3},\Theta_{4},\Theta_{5},\Theta_{6}\right) & = & S_{+-}^{+-}\left(\Theta_{4}-\Theta_{3}\right)F_{--+-++}\left(\Theta_{1},\Theta_{2},\Theta_{4},\Theta_{3},\Theta_{5},\Theta_{6}\right)\nonumber \\
 &  & \qquad+S_{+-}^{-+}\left(\Theta_{4}-\Theta_{3}\right)F_{---+++}\left(\Theta_{1},\Theta_{2},\Theta_{4},\Theta_{3},\Theta_{5},\Theta_{6}\right)\label{W6_1}.
\end{eqnarray}
Test results for the six-particle form factors are listed in Table 3.

\begin{table}\label{tab6}
\centering
\begin{tabular}{|c|c|c|}
  \hline
& LHS & RHS\\
\hline
  (\ref{W6_1}), $\xi=1.17$ &$-0.50782662-2.333030973i$ &$-0.50782660-2.33030977i$ \\
  (\ref{W6_1}), $\xi=0.34$ &$-0.3945330-0.3095434i$ &$-0.3945333-0.3095431i$ \\
  (\ref{kin6}), $\xi=1.17$ &$-2.84279-1.63925i$ & $-2.84263-1.63902i$\\
  (\ref{kin6}), $\xi=0.34$ &$-0.0151096-0.147483i$ &$-0.0151107-0.147442i$ \\
  \hline
\end{tabular}
\caption{Comparison of the LHS's and RHS's of the form factor axioms in the six-soliton case. 
For (\ref{W6_1}) the rapidities were chosen to be  $\Theta_{1}=2.1,\,\Theta_{2}=1.9,\,\Theta_{3}=6,\,\Theta_{4}=5.9,\,\Theta_{5}=1.2,\,\Theta_{6}=5.5+i\pi$, while for (\ref{kin6}), $\Theta_{1}=2.1,\,\Theta_{2}=1.9,\,\Theta_{3}=5.9,\,\Theta_{4}=1.2,\,\Theta_{5}=6,\,\Theta_{6}=5.90001+i\pi$ was taken. In all the tests $\frac{a}{\beta}=\frac{5}{4}$ was set.}
\end{table}

It interesting to note, that the use of multi-soliton form factors extends to the calculation of soliton-breather and breather-breather form factors by virtue of the bound state pole axiom \cite{Smirnov:1992}. In case of higher breather-breather form factors, using soliton-antisoliton form factors can be preferable: e.g. to calculate the $B_n$--$B_m$ form factor one needs to evaluate either the $(n+m)$--$B_1$- or the four-soliton form factors.

\section{Conclusions and outlook}

We established a method to obtain the multi-soliton form factors numerically in the (1+1)-dimensional sine-Gordon model. The form factors are known in terms of integral representations, whose domains of convergence were extended by analytical continuation. In order to do this we needed the asymptotic series of the $W$-function. Detailed formulae were only shown for the four-soliton form factors, however the number of treatable particles is not limited by the procedure. Test results obtained by the code provided for the four- and six-soliton form factors were shown.

Based on the formalism developed for finite volume form factors in \cite{PT1:2007,PT2:2008,Pmu:2008}, a program to investigate the sine-Gordon form factors is currently underway \cite{FeherTakacs}, which can now be extended to multi-soliton states \cite{FPT:2011}. The present formalism is also expected to be relevant to boundary form factors \cite{BPT:2006}, for which finite size corrections have been developed in \cite{KT:2007} and applied to sine-Gordon theory in \cite{LT:2011}. Future applications will also include the calculation of finite temperature correlation functions based on the formalism developed in \cite{Essler:2007,Essler:2009,Pozsgay}.

\section*{Acknowledgements}

The author is indebted to G\'abor Tak\'acs for a number of valuable discussions and for carefully reading the manuscript.

\bibliographystyle{utphys}
\bibliography{sgff}

\end{document}